\documentclass[conference]{IEEEtran}

\usepackage{cite}

\ifCLASSINFOpdf
   \usepackage[pdftex]{graphicx}
   \graphicspath{{../pdf/}{../jpeg/}}
   \DeclareGraphicsExtensions{.pdf,.jpeg,.png}
\else
  \usepackage[dvips]{graphicx}
   \graphicspath{{../eps/}}
   \DeclareGraphicsExtensions{.eps}
\fi

\usepackage[utf8]{inputenc}

\usepackage[table]{xcolor}      % Für die Farbigkeit
\definecolor{blue}{rgb}{0,0.29,0.6}  % 0, 74, 153
\definecolor{darkblue}{rgb}{0,0.216,0.447}  % 0, 55, 114
\definecolor{green}{rgb}{0.502,0.686,0.22}  % 128, 175, 56
\definecolor{darkgreen}{rgb}{0.373,0.514,0.165}  % 95, 131, 42
\definecolor{purple}{rgb}{0.573,0.188,0.498}  % 146, 48, 127
\definecolor{darkpurple}{rgb}{0.427,0.141,0.369}  % 109, 36, 94
\definecolor{beige}{rgb}{.898,.918,.867}
\definecolor{grey}{rgb}{.33, .33, .33}

\usepackage{listings}			% Quellcode-Listings
\begin{document}

\newcommand{\TODO}[1]{\textbf{\sffamily {TODO: #1}}}
%\renewcommand{\TODO}[1]{}

\title{Using Large-Scale Local and Cross-Location Experiments for Smart Grid System Validation}
%\title{Co-Simulation Concepts for Large-Scale Local and Cross-Location Experiments in the Energy Domain}

\author{\IEEEauthorblockN{Martin B\"uscher, Sebastian Lehnhoff, Sebastian Rohjans}
\IEEEauthorblockA{R\&D Division Energy\\OFFIS -- Institute for Information Technology\\
Oldenburg, Germany\\
Email: \{buescher, lehnhoff, rohjans\}@offis.de}
\and
\IEEEauthorblockN{Filip Andr\'en, Thomas Strasser}
\IEEEauthorblockA{Electric Energy Systems -- Energy Department\\AIT Austrian Institute of Technology\\ Vienna, Austria\\
Email: \{filip.andren, thomas.strasser\}@ait.ac.at}}


%Telephone: (800) 555--1212, Fax: (888) 555--1212}


\maketitle

\begin{abstract}
For robust testing of new technologies used in future, intelligent power and energy systems, realistic testing environments are needed. Due to the dimensions of a real-world environment a field-based installation is often not viable. More efficient instead of a local installation is to connect existing and highly sophisticated labs with different focus of specialization. Today’s experimental setups for the Smart Grid domain are very time-consuming solutions or specific implementations for a single project. To overcome this challenge, an innovative concept for a novel approach for large-scale co-simulation across locations (different labs) is presented in this paper.
\end{abstract}


\section{Introduction}\label{intro}
Historically, the power grid has been built hierarchically following a top-down architecture with generation mainly from centralized fossil power plants. The power flow in this topology is uni-directional from high to low voltage levels. Therefore, the operation of the power grid was feasible with less technical effort. Today's power grid infrastructure is changing compared to the top-down architecture. An important reason for this is the integration of an increasing number of Distributed Energy Resources (DER) like wind, Photovoltaic (PV), or Combined Heat and Power (CHP) plants \cite{Farhangi:2010}. Unlike the fossil power plants, most of the DER-based generators are not connected to the high voltage levels but to lower voltage levels. Due to this feed in phenomena like unstable frequencies, fluctuating voltage, and reverse load flows can be observed \cite{Lehnhoffrulz}. Furthermore, not only the topological changes have to be considered but also various uncertainties, e.g., caused by volatile DER generation -- that are not neglectable even with state-of-the-art prediction approaches -- have to be taken into account \cite{blank2014}. 

Additionally, due to changing framework conditions and technology developments like the liberalization of the energy markets, changing regulatory rules as well as the development of new components, the design and operation of the future electric energy system have to be restructured. Sophisticated component design methods, intelligent information and communication architectures, automation and control concepts as well as proper standards are necessary in order to manage the higher complexity of such intelligent power systems (i.e., Smart Grids) \cite{Farhangi:2010,IEA:2011}. Furthermore, besides the technical challenges economic, ecological but also social issues have to be addressed in Smart Grid research and innovation. 

\begin{figure}[htb]
	\centering
		\includegraphics[width=0.95\linewidth]{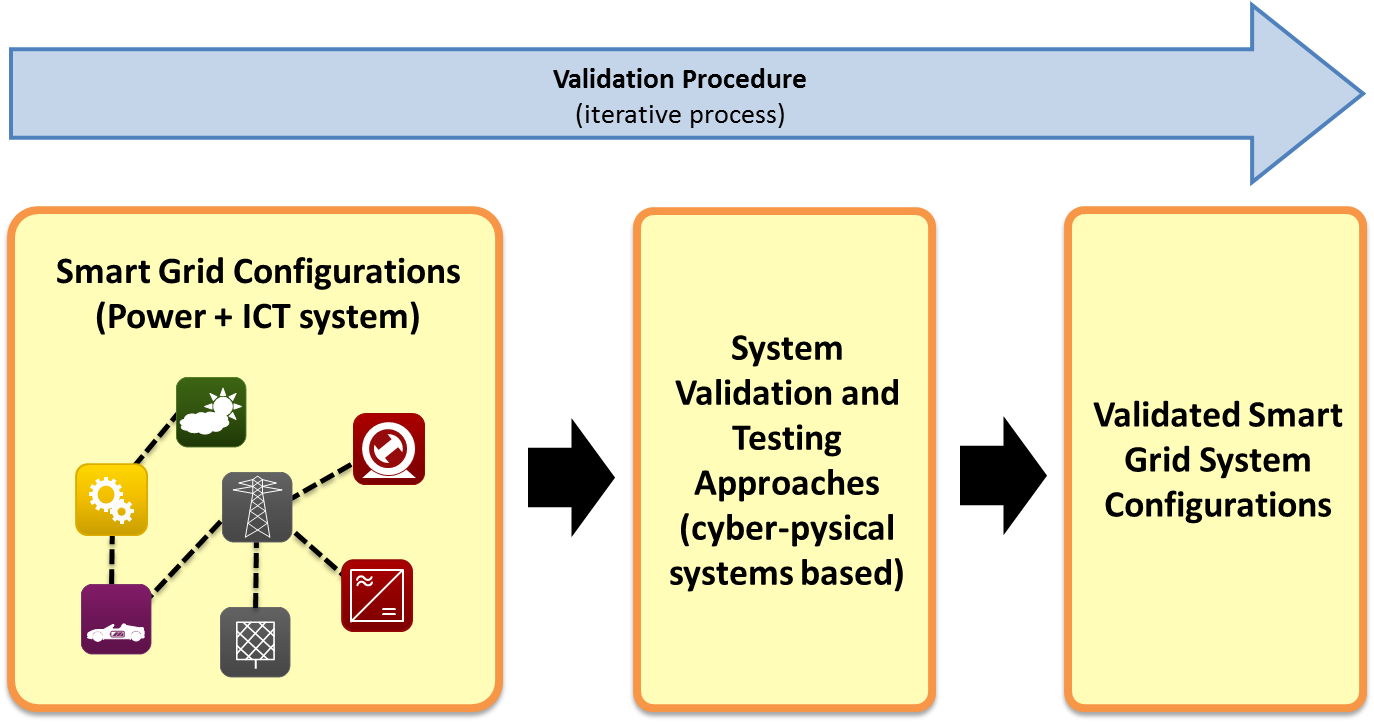}
	\caption{Overview of an integrated Smart Grid validation process.}
	\label{fig:ValidationProcess}
\end{figure}

Due to the increasing complexity of the overall energy system, a higher demand of automatized operational optimization becomes necessary \cite{finalsnord}. Therefore, a comprehensive validation of future Smart Grid components is mandatory in order to guarantee a reliable electricity supply. The complexity is based on a variety of non-linear interactions in future power and energy systems, so that small effects can have relevant impact \cite{Buscher2014a}. Testing only single Smart Grid components and thus ignoring their integration in the overall system is not sufficient because the complex interactions between the components are neglected. Hence, an integrated and holistic validation process (Fig.~\ref{fig:ValidationProcess}) for Smart Grid system configurations and not only for separated components is required.

This requires to find the suitable complexity of a validation environment. Because of the dimensions and the expected costs for a realistic testing environment, one single local validation environment in only one laboratory is often not feasible. However, it might be more efficient to connect already existing and established laboratory environments with different focus of specialization and facilities to build up large-scale co-simulation-based environments. Such a setup would reflect the large-scale real-world setting as close as possible \cite{Faschang:2015}. State of the art approaches are limited to very time-consuming solutions or specific implementations for a single project. Therefore, testing algorithms or systems in a large-scale co-simulation platform with a realistic model of the real power grid -- with all its components and characteristics -- is only possible with great effort and expert knowledge.

The main goal of this paper therefore is to analyze the needs for a proper Smart Grid validation and testing infrastructure as well as to introduce a methodology to model and execute local but also coupled cross-location experiments of distributed laboratory environments. 

The remainder of this paper is structured as follows: In Section~\ref{needsForTestingInfrastructure} the needs for a holistic, cyber-physical based Smart Grid testing infrastructure are elaborated. The following Section~\ref{objectives} discusses requirements for a validation/testing system and in Section~\ref{concept} the composed system is presented. A realized validation example is demonstrated in Section~\ref{ValidationExample}. Finally, Section~\ref{conclusions} summarizes the discussion in this paper and provides an outlook of planned further work.
\section{Needs for a Smart Grid Validation and Testing Infrastructure}\label{needsForTestingInfrastructure}
\subsection{A Cyber-Physical Systems Approach}
Due to the considerable higher complexity of Smart Grid configurations -- which can be considered as cyber-physical systems -- it is expected that their validation will play a major role in future technology developments. Now, as the first demonstration projects for Smart Grid technologies were successfully finished, the probability that the key findings and achieved results will be integrated in new and/or modified products, solutions and services of manufacturers from the power system domain is quite high. 

Up to now no integrated approach for analyzing and evaluating Smart Grid configurations addressing the physical power system, as well as the information, communication and automation/control architecture is available. In order to guarantee a sustainable and secure supply of electricity in a Smart Grid system with considerable higher complexity as well as to support the expected forthcoming large-scale roll out of new technologies, a proper validation and testing infrastructure for Smart Grids is necessary as pointed out in the introduction. Such an infrastructure has to support system analysis, evaluation and testing issue. Furthermore, it would foster future innovations and technical developments in the field. In summary, the following open issues need to be addressed for the validation of Smart Grid concepts and approaches \cite{EEGI:2010,ETP:2012,IEA:2013}:

\begin{itemize}
	\item A cyber-physical (multi-domain) approach for analyzing and validating Smart Grids on the system level is missing today; existing methods are mainly focusing on the component level. Today, system integration topics including analysis and evaluation are not addressed in a holistic manner so far. 
	\item A holistic validation framework (incl. analysis and evaluation criteria) and the corresponding research infrastructure with proper methods and tools needs to be developed. 
	\item Harmonized and possibly standardized evaluation procedures need to be developed.
	\item Well-educated professionals, engineers and researchers understanding Smart Grid configurations in a cyber-physical manner need to be trained on a broad scale.
\end{itemize}

An early testing of different setups and configurations (incl. control and optimization algorithms) can already be achieved using simulation approaches (multi-domain simulation, co-simulation, software-in-the-loop simulation) \cite{Strasser:2014}. However, pure software simulation might be not enough anymore if software and/or hardware prototypes have to be evaluated. In such a case a proper laboratory infrastructure is necessary representing a near real-world environment. 

Since a real power system (e.g., transmission or distribution grid and its components) cannot be represented in a laboratory setup available power system/Smart Grid labs are  always covering a specific part of the real system (DER component testing infrastructure, ICT/automation systems lab, etc.). Due to the higher complexity of Smart Grid systems as pointed out above, an online coupling of different, available laboratories have to be considered in order to represent a more realistic near real-world setup. In the following sections representative examples of available Smart Grid-related labs are presented to underpin the above statements.
\subsection{Examples for Existing Laboratory Infrastructures}\label{existingLabInfrastructure}
Examples for existing Smart Grid related laboratories in Europe are provided by the DERlab\footnote{http://der-lab.net/derlabsearch/} association. The corresponding database shows details about the features of each of the different available laboratories and the available equipment which are typically used in various research, development and demonstration projects. Nonetheless, in the following two representative examples of Smart Grid related laboratories are provided; the AIT SmartEST lab representing a DER device validation and system integration testing infrastructure and the Smart Grid ICT/automation-related OFFIS SESA-Lab. 

\subsubsection{SmartEST Lab}
The AIT Smart Electricity Systems and Technologies (SmartEST) laboratory environment provides a multi-functional research, validation and testing infrastructure allowing to analyze the behavior but also the interactions between power system components -- especially inverter-based DER components -- and the power grid under realistic near real-world situations \cite{Bruendlinger:2015}. In general, the lab includes 3 configurable Low Voltage (LV) grids (3-phase), programmable, high-bandwidth power grid emulators as well as powerful Photovoltaic (PV) array simulators. An environmental test chamber for emulating different environmental conditions is also available. This allows validating and testing of inverter-based DERs at full power under extreme temperature and humidity conditions but also their interaction under different power grid configurations. Potential candidates for testing range from inverters, storage units, grid controllers and CHP units through to charging stations for electric vehicles in the power range of few kVA up to 1 MVA. 

Additionally, SmartEST allows the real-time simulation (via OPAL-RT's digital real-time simulator) of complex power grids incl. their components as well as the coupling of this virtual environment with the laboratory grids. Therefore, such kind of Hardware-in-the-Loop (HIL) setup provides the possibility to integrate real power system components into a virtual grid environment and to be tested under realistic conditions in interaction with the grid. Besides the HIL-based integration of power system components also ICT approaches, concepts and developments can be integrated into the whole setup allowing a comprehensive analysis of Smart Grid related topics. 

Furthermore, this laboratory provides also the possibility of online-coupling via Internet with other Smart Grid laboratories (e.g., ICT/automation) for more complex system studies and validation tasks.

\subsubsection{SESA-Lab}
In order to integrate and functionally combine software models of various quality, precision and model representation as well as integrating hardware components the University of Oldenburg and the OFFIS -- Institute for Information Technology have set up the Smart Energy Simulation and Automation Laboratory (SESA-Lab). The lab provides a platform for testing Smart Grid components in a realistic environment with the focus on the communication between the components. For this purpose the SESA-Lab is located in a closed local network which is reachable by a Virtual Private Network (VPN) for authorized users over a VPN Gateway. Furthermore, it is possible to divide the network into virtual sub-networks via VLAN to allow multiple experiments with a separated communication at the same time. To enable real-time communication between the components an Industrial Ethernet network (i.e., EtherCAT-based) is also available. To fulfill future requirements on communication the SESA-Lab is build highly flexible. All communication based on the Ethernet Standard IEEE 802.3 can be deployed easily.

For testing huge scenarios in the SESA-Lab different hardware components are available. For components with smaller tasks there are Programmable Logical Controllers (PLC) available, more demanding tasks can use industry PCs. Also available is an OPAL-RT eMEGAsim digital real-time simulator providing digital and analogue I/O interface capabilities. Therefore, physical systems can be simulated with high resolution in HIL experiments. A realistic validation of Smart Grids configurations is only possible if there are a large number of real components in the lab. Although the SESA-Lab provides some components, their number will be too small for huge realistic simulations. Therefore, a simulation server with the simulation framework mosaik (see Section~\ref{OrchestrationAndExecution}) is available to simulate all components which can't be represented in hardware. Even if the SESA-Lab is a suitable validation infrastructure for many scenarios using real hardware components like transformers, converter, photovoltaic systems, and other systems from the field of the power electronics is not in its scope. A cooperation with other labs -- which are specialized on this area -- is necessary for such kind of studies. 
\subsection{Coupling Lab Infrastructures}\label{couplingLabInfrastructure}
The above examples of state of the art laboratories (SmartEST Lab and SESA-Lab) clearly show the different specialization in the field of Smart Grid validation. As a result, it is clear that large-scale validation scenarios with many various technologies cannot be done in one local lab. One solution to this might be a lab with all necessary components and technologies inside. In principle, building such a lab is technically possible but operation and costs will be compelling reasons against this solution. This is due to the fact that all existing labs have own specifications and components and therefore experts for these components to guarantee a smooth operation as wall as the technical and theoretical competence in this field. Furthermore, today several labs with different components and features are already available. Therefore, to build up a single lab is financially and organizationally not feasible. 

\begin{figure}[htb]
	\centering
		\includegraphics[width=0.95\linewidth]{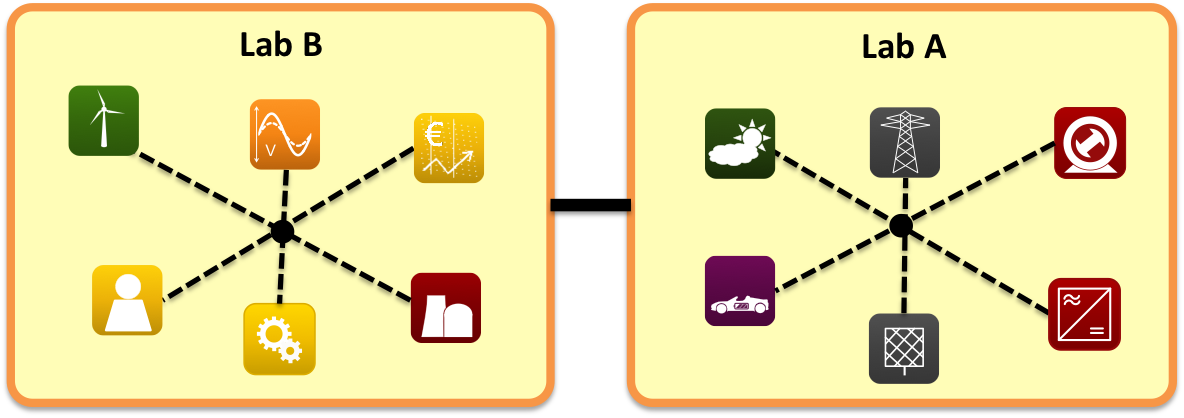}
	\caption{Coupling Lab infrastructures.}
	\label{fig:CouplingLabInfrastructures}
\end{figure}

To overcome this gap coupling existing lab infrastructures is an appropriate solution. In Fig.~\ref{fig:CouplingLabInfrastructures} an example for the coupling of two labs is conceptionally shown. Depending on the validation scenario two ore more labs have to be coupled to provide all necessary components. The coupling as well as the communication between the components in the different labs is not a trivial matter -- in the following, concepts for this will be presented and discussed.
\section{Setting up a Coupled Infrastructure}\label{objectives}
The previous section has exposed the need for a Smart Grid testing infrastructure and exemplary showed two existing laboratory infrastructures with different equipment and possibilities for Smart Grid validation. Now an example for setting up a coupled infrastructure for cross-laboratory experiments will be depicted. Today’s labs contain components for simulations, which can be initially used only in local experiments. The main reason for this is the missing communication beyond lab boundaries. Furthermore, concepts for modeling large-scale cross-location experiments are needed to facilitate planning the scenario and to conduct analysis in case of a fault. To overcome this gap the following points are of specific interest:

\begin{itemize}
\item modeling of the experiments,
\item local orchestration,
\item cross-location orchestration, and
\item coordinated execution.
\end{itemize}

These points will be depicted in the following sections.
\subsection{Modelling of Experiments}\label{modelling}
Before running and building up a Smart Grid validation scenario modeling –- especially for complex setups -– is an essential part of the preparation. Modeling the experiment has to be conducted on different layers, e.g., architecture, communication flow, component properties to describe the whole scenario. Although the different kinds of modeling tools allow a manual implementation and configuration of the systems, for large-scale scenarios this is not sufficient. Only with machine readable models an efficient automatic configuration is possible.
\subsection{Local Orchestration}\label{localOrchestrartion}
\begin{figure}[htp]
	\centering
		\includegraphics[width=0.75\linewidth]{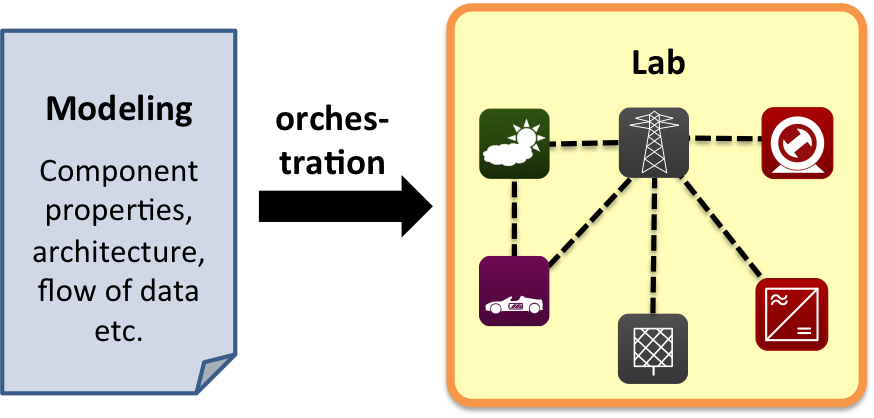}
	\caption{Local orchestration of lab components.}
	\label{fig:LocalOrchestrationLabComponents}
\end{figure}

The first step on the path to an automatic configuration across different locations is the local orchestration. Local orchestration means to use a machine readable model of the deliverable scenario, in order to configure all components in a local lab automatically. In Fig.~\ref{fig:LocalOrchestrationLabComponents} the local orchestration is depicted in a schematic illustration. The components in the presented lab can have different kinds of network interfaces and/or communication protocols. The challenge is to consolidate all of the components to enable a transparent communication, e.g., with network and/or protocol adapters.

\subsection{Cross-Location Orchestration}\label{cossLocationOrchestration}
\begin{figure}[htp]
	\centering
		\includegraphics[width=0.75\linewidth]{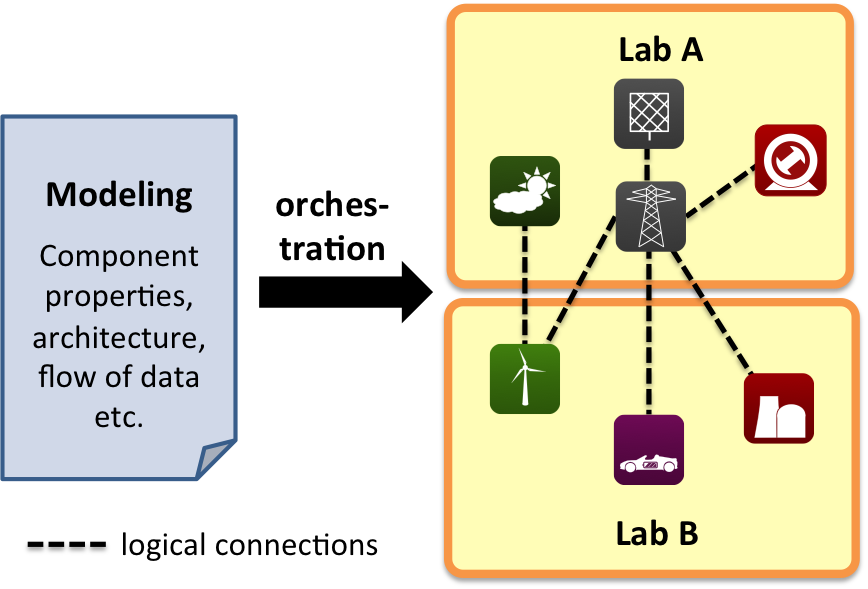}
	\caption{Orchestration of lab components across locations.}
	\label{fig:OrchestrationLabComponentsAcrossLocations}
\end{figure}

On the basis of the local orchestration the proposed concept has to be expanded to enable the orchestration across locations (involvement of different, specific laboratories). This is depicted in Fig.~\ref{fig:OrchestrationLabComponentsAcrossLocations}.

The first step, to build a machine readable model of the system, is the same as described in Section~\ref{modellingOfExperiments}. Thus, for the system user (who specifies the model) it is no difference to run the scenario in a local lab or in a configuration setup with two or more labs.

For getting logical connections (for example over the Internet) between the components in different labs, the orchestration process has to be adapted.
\subsection{Coordinated Execution}\label{coordinatedExecution}
\begin{figure}[htp]
	\centering
		\includegraphics[width=0.70\linewidth]{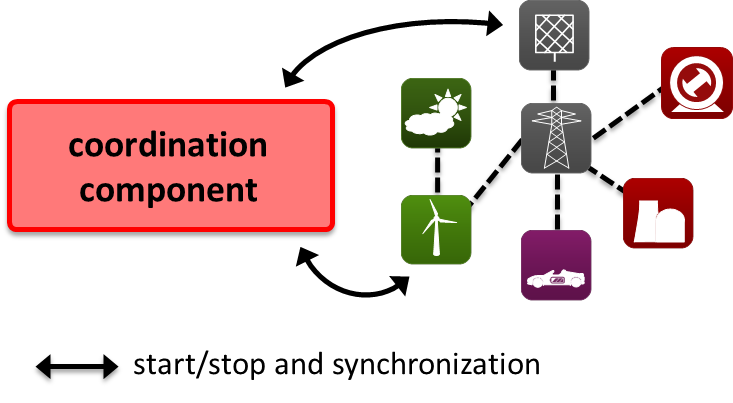}
	\caption{Coordinated execution of lab components.}
	\label{fig:CoordinatedExecutionOfLabComponents}
\end{figure}

After modeling and orchestration of the Smart Grid validation scenario as a local lab or a across-location lab experiment, the validation has to be started. For this purpose a component -- as shown in Fig.~\ref{fig:CoordinatedExecutionOfLabComponents} -- is necessary to coordinate the execution. One task of this component to start all components in the validation scenario at the same time as well as to inform all components to stop after the validation is finished or should be canceled. Furthermore the synchronization of all involved components -- in case of different temporal resolutions -- during the validation is the main part of the coordination component.  

\section{Proposed Basic Concept}\label{concept}
In the previous section the objectives for the concept were discussed. Now, the requirements to build up such a system will be identified and compared with the state of the art. Also a first potential concept is presented and discussed.
\subsection{Communication}\label{communication}
\begin{figure}[htb]
	\centering
		\includegraphics[width=0.75\linewidth]{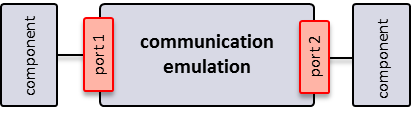}
	\caption{Communication simulation.}
	\label{fig:CommunicationSimulation}
\end{figure}

For realistic simulations in the power and energy domain (especially for Smart Grid related applications with remote control/monitoring, etc.) it is necessary to simulate the systems and algorithms in an experiment, but also considering the realistic communication between the components is an important factor. This is justified by the fact that communications in the real world are not without errors and also have bandwidth limitations. Therefore, the communication between lab components has to exhibit the same behaviors as the communication in the real world. To achieve such communication simulations, real-time emulation of communication networks and systems \cite{NS3Wiki2014, Wehrle2010, Necker2006} as shown in Fig.~\ref{fig:CommunicationSimulation} have to be deployed. With a real-time emulation of such networks it is than possible to establish a realistic communication (e.g., mobile networks) between two real (hardware) components \cite{Lin2011,Strasser:2014}. This measure contributes testing the system behavior with defective communication and not only with a idealistic communication that usually is available in lab environments. This is important for the validation of Smart Grids because system stability and robustness in case of expected communication errors or bandwidth limitations is crucial for a high threat infrastructure such as the energy system.

Furthermore, not only the local communication is relevant for Smart Grid validation -- the behavior of cross-location communication between labs in huge validation scenarios as well as the mutual interference of parallel experiments has to be considered. Cross-location communications represent a particular challenge for the validation because of the interferences and bandwidth limitations itself. With suitable concepts it has to be guaranteed that the cross-location communication does not have any influence on the communication of the validated components. It must furthermore be assumed that lab environments have a large number of components for different applications. Therefore, it should be considered that only a part of the available components will be used in a particular scenario. For this reason concepts for running different validations in parallel to achieve the best possible utilization of the expensive lab components have to be developed.
\subsection{Modelling of Experiments}\label{modellingOfExperiments}
%\textcolor{red}{\textbf{ToDo (Thomas): Unterabschnitt zu Modellierung mit MDE, Formale Test-Case Beschreibung.}}
%
%\textcolor{red}{\textbf{ToDo (Martin und/oder Sebastian R.): Unterabschnitt SGAM als Modellierungswerkzeug sowie ggf. OPC UA}}
%
%\textcolor{red}{\textbf{ToDo (Martin): Abschnitt überarbeiten und bestehende Teile zusammenführen}}
A proper Smart Grid scenario description method is necessary in order to model an experiment. Such a description has to cover the test/validation case (i.e., ``What needs to be evaluated?'') but also what is available from the laboratory side (i.e., ``What can be provided/tested?''). Furthermore, it should also provide the possibility to (semi-)automatically derive lab configurations out of it. Up to now no suitable test description language exists for the power and the energy domain. It is subject for further research. In the following some ideas and available approaches from other domains are presented which provide the basis for such a required Smart Grid validation description method and corresponding language.

As a foundation, the Smart Grid Architecture Model (SGAM) developed under EU Mandate M/490 will be taken into consideration. The SGAM defines a structured approach for Smart Grid architecture development. Furthermore, the IEC Publicly Available Specification (PAS) 62559 template (adopted by IEC~TC8/WG5) and method for use case development will be employed to gather relevant information. A key focus of this task will be on extending this use case description, based on the efforts in IEC~TC8/WG5, with coherently structured quantitative quality and test criteria for the proper valuation of uncertainty and significance of experimental setups. 

As already illustrated in Section~\ref{modelling} a machine readable model of all relevant layers is important to allow for an automated orchestration. Common standards for describing components -– respectively their interfaces –- like IEC~61850, IEC~61499, UML, Common Information Model (IEC~61970/61968; CIM \cite{lackingcim}), OPC~UA etc. are partially machine readable. The same applies for the architecture and data flow models like UML or the SGAM \cite{CEN-CENELEC_SGAM2012, daenekas2014}. Furthermore, there is no general model for all layers to be considered. For this, today’s modeling tools are not sufficient. Hence, suitable adaptions are necessary.

Furthermore, an automatic transformation of information and data of testing scenarios/validation cases represented in the different SGAM layers into machine readable models as pointed out above, is necessary. Model-Driven Design (MDD) and the corresponding Model-Driven Architecture (MDA) concept \cite{Mellor:2004,Siegel:2001}, well known form the computer science domain, are potential candidates for this issue. They mainly focus on the development of domain specific software models using a Platform Independent Model (PIM) for representing application software, the description of the corresponding execution platform using a Platform Specific Model (PSM) as well as the mapping of both models. This approach also allows platform specific code generation introducing the so-called Implementation Specific Model (ISM). The ideas and concepts behind can be used to develop a modeling framework describing validation scenarios for the Smart Grid domain. 

Summarizing, for the modeling of experiments a proper description method and a corresponding domain-specific language have to be developed. Together with a corresponding tool framework (potential candidates have been briefly sketched above) a powerful environment for the modeling and configuration of cross-location experiments can be setup. 
\subsection{Orchestration and Execution}\label{OrchestrationAndExecution}
The final step after the modeling phase is the automated orchestration of the scenario (see Sections~\ref{localOrchestrartion} and \ref{cossLocationOrchestration}). Mosaik \cite{isgt} for time discrete simulations and the Simulation Message Bus (SMB) concept \cite{Faschang2013,Mosshammer2013} for time continuous simulations, are promising approaches for orchestrating components or software tools:

\begin{itemize}
	\item{\textit{Mosaik:}} Mosaik\footnote{http://mosaik.offis.de} is a flexible open source Smart Grid co-simulation framework written in Python. It enables to combine easily existing simulation models for creating large-scale scenarios. The scenarios in mosaik can have thousands of simulated entities to accomplish a high level of detail. All simulators that should be used in a mosaik simulation have to use the mosaik Application Programming Interface (API) which is available for different programming languages (e.g., Python, Java, C\#). After integrating the mosaik API in the used simulators, the scenario description has to be made with mosaik. The event-based execution coordination of the connected models is afterwards done by mosaik.
\item{\textit{Simulation Message Bus:}} For coupling components in the prototyping or validation process the SMB provides a suitable platform. With the SMB all connected components can communicate over a stand-alone server component that routes messages like a network switch. Every component has to implement a connector, which translates the used protocol (OPC~UA, IEC~61850, etc.) to the SMB protocol. Therefore, real hardware components can be integrated easily with an appropriate connector. The advantage of the SMB is the continuous communication for the whole development process. For the purpose of building up a Smart Grid validation of continuous models or real hardware components the SMB is well suited. 
\end{itemize} 

The problem of the presented approaches is that they are suitable either for discrete simulations or for continuous simulations. To allow realistic simulations the full bandwidth of models (from time discrete models to continuous real-time models) has to be combined in one approach. Moreover, current approaches are not able to manage components in different labs as well as allowing fully automated orchestration of scenarios with different component protocols and interfaces. Therefore, a novel concept has to be found to combine the benefits of both (mosaik and SMB) systems and give furthermore the possibility for cross-location orchestrations. This issue is subject for further research.
\section{Validation Example}\label{ValidationExample}
\begin{figure}[htb]
	\centering
		\includegraphics[width=0.9\linewidth]{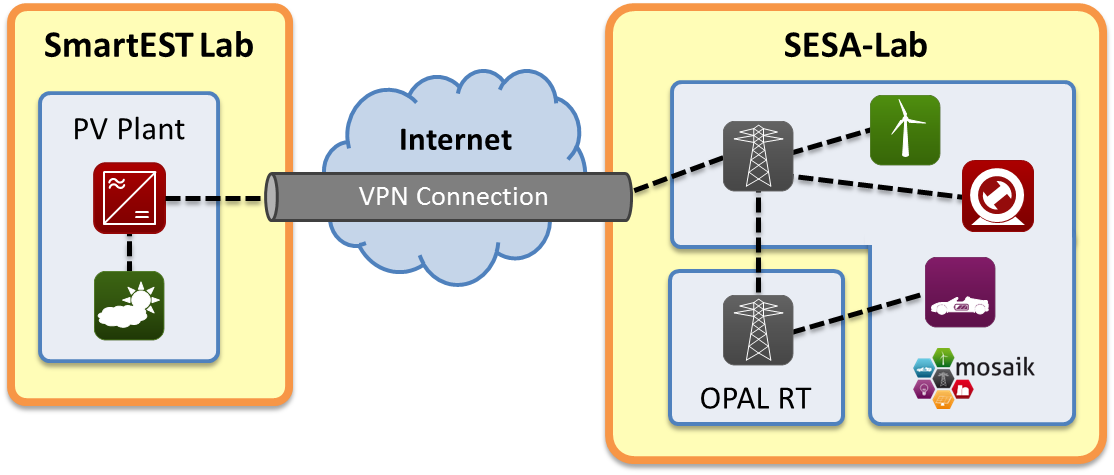}
	\caption{Linkage between SmartEST Lab and SESA-Lab.}
	\label{fig:linkageSmartEstAndSesaLab}
\end{figure}

After presenting the concept for a large-scale local and cross-location Smart Grid validation, a representative example \cite{Buscher2014a} has been chosen to demonstrate that coupling labs over huge distances is possible and useful. In this example of a Smart Grid system's validation the co-simulation framework mosaik, a real-time simulator (OPAL-RT) and an real PV inverter is used (see Fig. \ref{fig:linkageSmartEstAndSesaLab}). In this co-simulation study the behavior of various PV systems connected in a low-voltage power distribution grid was analyzed. 

The coordination of the simulation is managed by mosaik in the SESA-Lab. Mosaik starts a network calculation and annotates simulated consumers and prosumers to the notes of the simulated network. Furthermore, the simulated network is extended by another network calculated in the OPAL-RT real-time simulator. Hereby it is possible to analyze a small part of the real-time calculated network with high accuracy. As already noted in Section~\ref{needsForTestingInfrastructure} it is not feasible to use real power electronic components from the field in the SESA-Lab. Therefore, the SESA-Lab was coupled with a real PV inverter which is available in the SmartEST Lab. By linking the two labs together the simulation could be done better and more accurate than using only a simulation model for the PV system. 

This small example shows how useful cross-location experiments between labs -- with its various application potentials -- for huge Smart Grid simulations can be. Furthermore, it became clearly evident that coupling components from different labs is a complex process because of the different technologies, communication protocols, network infrastructures, and necessary domain expertise. Only by simplifying this process in the presented manner cross-location experiments for Smart Grid system validation is practicable.
\section{Summary and Conclusions}\label{conclusions}
In this contribution a basic concept for large-scale co-simulation across multiple labs was motivated. The objectives for such a system were identified and the requirements to achieve the objectives were described. Furthermore, it was shown, that different approaches and modeling tools exist that meet the objectives partly. However, a solution fulfilling all objectives is not available.
Future work will mainly focus on the machine readable experiment modeling based on common models and standards. Furthermore, detailed concepts for local and communication between labs will be developed.

% \fi
\IEEEpeerreviewmaketitle

%\section*{Acknowledgment}
%\textcolor{red}{\textbf{ToDo: ggf. Einfügen}}

%\IEEEtriggeratref{7}


\bibliographystyle{IEEEtran}
\bibliography{ref}

\begin{thebibliography}{10}
\providecommand{\url}[1]{#1}
\csname url@samestyle\endcsname
\providecommand{\newblock}{\relax}
\providecommand{\bibinfo}[2]{#2}
\providecommand{\BIBentrySTDinterwordspacing}{\spaceskip=0pt\relax}
\providecommand{\BIBentryALTinterwordstretchfactor}{4}
\providecommand{\BIBentryALTinterwordspacing}{\spaceskip=\fontdimen2\font plus
\BIBentryALTinterwordstretchfactor\fontdimen3\font minus
  \fontdimen4\font\relax}
\providecommand{\BIBforeignlanguage}[2]{{%
\expandafter\ifx\csname l@#1\endcsname\relax
\typeout{** WARNING: IEEEtran.bst: No hyphenation pattern has been}%
\typeout{** loaded for the language `#1'. Using the pattern for}%
\typeout{** the default language instead.}%
\else
\language=\csname l@#1\endcsname
\fi
#2}}
\providecommand{\BIBdecl}{\relax}
\BIBdecl

\bibitem{Farhangi:2010}
H.~Farhangi, ``{The path of the smart grid},'' \emph{IEEE Power and Energy
  Magazine}, vol.~8, no.~1, pp. 18--28, 2010.

\bibitem{Lehnhoffrulz}
S.~Lehnhoff, T.~Klingenberg, M.~Blank, M.~Calabria, and W.~Schumacher,
  ``{Distributed Coalitions for Reliable and Stable Provision of Frequency
  Response Reserve},'' in \emph{IEEE International Workshop on Intelligent
  Energy Systems}, 2013, pp. 11--18.

\bibitem{blank2014}
M.~Blank and S.~Lehnhoff, ``{Correlations in Reliability Assessment of
  Agent-Based Ancillary-Service Coalitions},'' in \emph{Power Systems
  Computation Conference (PSCC)}, 2014.

\bibitem{IEA:2011}
\BIBentryALTinterwordspacing
``{Technology Roadmap Smart Grids},'' {International Energy Agency (IEA)},
  Tech. Rep., 2011. [Online]. Available: \url{http://www.iea.org}
\BIBentrySTDinterwordspacing

\bibitem{finalsnord}
L.~Hofmann and M.~Sonnenschein, Eds., \emph{{Smart Nord - Final Report}}, 2015.

\bibitem{Buscher2014a}
M.~Buscher, A.~Claassen, M.~Kube, S.~Lehnhoff, K.~Piech, S.~Rohjans,
  S.~Scherfke, C.~Steinbrink, J.~Velasquez, F.~Tempez, and Y.~Bouzid,
  ``{Integrated Smart Grid simulations for generic automation architectures
  with RT-LAB and mosaik},'' in \emph{2014 IEEE International Conference on
  Smart Grid Communications (SmartGridComm)}, 2014, pp. 194--199.

\bibitem{Faschang:2015}
M.~Faschang, F.~Kupzog, E.~Widl, S.~Rohjans, and S.~Lehnhoff, ``{Requirements
  for Real-Time Hardware Integration into Cyber-Physical Energy System
  Simulation},'' in \emph{2015 IEEE Workshop on Modeling and Simulation of
  Cyber-Physical Energy Systems}, 2015.

\bibitem{EEGI:2010}
\BIBentryALTinterwordspacing
``{Roadmap 2010-18 and Detailed Implementation Plan 2010-12},'' {European
  Electricity Grid Initiative (EEGI)}, Tech. Rep., 2010. [Online]. Available:
  \url{http://www.smartgrids.eu}
\BIBentrySTDinterwordspacing

\bibitem{ETP:2012}
\BIBentryALTinterwordspacing
``{SmartGrids SRA 2035 – Strategic Research Agenda, Update of the Smart Grids
  SRA 2007 for the needs by the year 2035},'' {European Technology Platform
  Smart Grids}, Tech. Rep., 2012. [Online]. Available:
  \url{http://www.smartgrids.eu}
\BIBentrySTDinterwordspacing

\bibitem{IEA:2013}
\BIBentryALTinterwordspacing
``{International Smart Grid Action Network (ISGAN) – Annex 5: Smart Grid
  International Research Facility Network (SIRFN)},'' {International Energy
  Agency (IEA)}, Tech. Rep., 2013. [Online]. Available:
  \url{http://www.iea.org}
\BIBentrySTDinterwordspacing

\bibitem{Strasser:2014}
T.~Strasser, M.~Stifter, F.~Andren, and P.~Palensky, ``Co-simulation training
  platform for smart grids,'' \emph{IEEE Transactions on Power Systems},
  vol.~29, no.~4, pp. 1989--1997, July 2014.

\bibitem{Bruendlinger:2015}
R.~Br\"{n}dlinger, T.~Strasser, G.~Lauss, A.~Hoke, S.~Chakraborty, G.~Martin,
  B.~Kroposki, J.~Johnson, and E.~de~Jong, ``Lab tests: Verifying that smart
  grid power converters are truly smart,'' \emph{IEEE Power and Energy
  Magazine}, vol.~13, no.~2, pp. 30--42, March 2015.

\bibitem{NS3Wiki2014}
{NS-3 Wiki}, ``{Howto make ns-3 interact with the real world}.''

\bibitem{Wehrle2010}
K.~Wehrle, G.~Mesut, and J.~Gross, Eds., \emph{{Modeling and Tools for Network
  Simulation}}.\hskip 1em plus 0.5em minus 0.4em\relax Berlin, Heidelberg:
  Springer Verlag, 2010.

\bibitem{Necker2006}
M.~Necker and U.~Reiser, \emph{{IKR Emulation Library 1.0 User Guide}}.\hskip
  1em plus 0.5em minus 0.4em\relax Stuttgart: Institut f\"{u}r
  Kommunikationsnetze und Rechnersysteme, Universit\"{a}t Stuttgart, 2006.

\bibitem{Lin2011}
H.~Lin, S.~Sambamoorthy, S.~Shukla, J.~Thorp, and L.~Mili, ``{Power system and
  communication network co-simulation for smart grid applications},'' in
  \emph{2011 IEEE PES Innovative Smart Grid Technologies (ISGT)}, 2011, pp.
  1--6.

\bibitem{lackingcim}
M.~Uslar, S.~Rohjans, M.~Specht, and J.~Gonzalez, ``{What is the CIM
  lacking?}'' in \emph{IEEE ISGT Europe}, 2010.

\bibitem{CEN-CENELEC_SGAM2012}
{CEN-CENELEC-ETSI Smart Grid Coordination Group}, \emph{{Smart Grid Reference
  Architecture}}, 2012.

\bibitem{daenekas2014}
C.~D\"{a}nekas, C.~Neureiter, S.~Rohjans, M.~Uslar, and D.~Engel, ``{Towards a
  Model-Driven-Architecture Process for Smart Grid Projects},'' in
  \emph{Digital enterprise design \& management}, 2014, pp. 47--58.

\bibitem{Mellor:2004}
S.~J. Mellor, S.~Kendall, A.~Uhl, and D.~Weise, \emph{{MDA Distilled}}.\hskip
  1em plus 0.5em minus 0.4em\relax Redwood City, CA, USA: Addison Wesley
  Longman Publishing Co., Inc., 2004.

\bibitem{Siegel:2001}
J.~Siegel, ``{Developing in OMG's New Model-Driven Architecture},''
  \emph{Management}, 2001.

\bibitem{isgt}
S.~Rohjans, S.~Lehnhoff, S.~Sch\"{u}tte, S.~Scherfke, and S.~Hussain, ``{mosaik
  -- A modular Platform for the Evaluation of Agent--Based Smart Grid
  Control},'' in \emph{4th IEEE PES International Conference and Exhibition on
  Innovative Smart Grid Technologies (ISGT Europe)}, 2013.

\bibitem{Faschang2013}
M.~Faschang, F.~Kupzog, R.~Mosshammer, and A.~Einfalt, ``{Rapid control
  prototyping platform for networked smart grid systems},'' \emph{IECON
  Proceedings (Industrial Electronics Conference)}, pp. 8172--8176, 2013.

\bibitem{Mosshammer2013}
R.~Mosshammer, F.~Kupzog, M.~Faschang, and M.~Stifter, ``{Loose coupling
  architecture for co-simulation of heterogeneous components},'' \emph{IECON
  Proceedings (Industrial Electronics Conference)}, pp. 7570--7575, 2013.

\end{thebibliography}

\end{document}